**Title: In-store epidemic behavior: scale development and validation**

**Andrzej SZYMKOWIAK[1*], Piotr KULAWIK[2], Kishokanth JEGANATHAN[1], Paulina GUZIK[2]**

[1] Department of Commerce and Marketing, Institute of Marketing, Poznań University of Economics and Business, ul. Niepodległosci 10, 61-875 Poznań, Poland

[2] Department of Animal Products Technology, Faculty of Food Technology, University of Agriculture in Kraków, ul. Balicka 122, 30-149 Kraków, Poland

[*]**Correspondence should be addressed to:**

Andrzej Szymkowiak (Ph.D.)

Department of Commerce and Marketing, Institute of Marketing, Poznań University of Economics and Business, ul. Niepodległosci 10, 61-875 Poznań, Poland, + 48 508133038, andrzej.szymkowiak@ue.poznan.pl

**Abstract:** Epidemics of infectious diseases have accompanied humans for a long time and, depending on the scale, cause various undesirable social and economic consequences. During the ongoing COVID-19 pandemic, governments of many countries impose restrictions to inhibit spreading of infection. Isolation and limiting interpersonal contacts are particularly recommended actions. Adhering to the rule of isolation may involve restrictions in freedom during daily activities, such as shopping. The aim of the study was to develop a scale of in-store pandemic behavior. The whole process involved 3 stages: qualitative inquiry, scale purification and scale validation, which were based on 3 studies: 1 qualitative (20 in-depth interviews) 2 two quantitative (373 and 584 respondents, respectively), and allowed to identify 8 factors. Following, a theoretical model was created to investigate the impact of in-store infection threat on identified variables. All identified factors significantly correlated with the in-store infection threat which reiterates the importance of providing information revealing the true scale of the pandemic and not leaving space for individuals to create subjective probability judgments. The developed scale can help counteract disinformation and assess consumer behavior compliance and understanding of the official recommendations imposed by governments, enabling more efficient educational efforts.

**Keywords:** COVID-19; in-store behavior, epidemic, threat perception

# Introduction

Viral infections and epidemics have plagued humanity for generations, with many researchers indicating their occurrence as inevitable (Funk, Salathé, & Jansen, 2010; Kuiken, Fouchier, Rimmelzwaan, & Osterhaus, 2003). Since 1940, approximately 400 emerging infectious diseases have been identified, with most of them being zoonotic (Morse et al., 2012). These, in turn, cause many undesirable effects such as an increased mortality rate and economic impact on society (Heymann, 2005; Salathé et al., 2010).

The majority of zoonotic diseases require direct contact with an infected animal as was the case with malaria, yellow fever or Zika, which are transmitted through mosquito bites (Abeku et al., 2004; Ahmed Ali, Nyla, Mashael, Salvatore, & Mohammed, 2016; Briand et al., 2009; Carey, Wang, Su, Zwiebel, & Carlson, 2010; Ferguson et al., 2016; Lucey & Gostin, 2016; Wanjala, Waitumbi, Zhou, & Githeko, 2011; Wasserman, Tambyah, & Lim, 2016), or avian flu (H5N1), in which the main vectors are poultry and wild birds (Lewis, 2006; Peiris, De Jong, & Guan, 2007; Woo, Lau, & Yuen, 2006). In the absence of a vaccine, the best means of prevention against such diseases is avoidance and protection against potentially infected individuals (Craft, 2015; Yousaf et al., 2012). Another possible transmission route is through the consumption of infected feces, mainly through oral means. Examples of such transmitted diseases include rotavirus, norovirus or hepatitis A. The key to protecting oneself, apart from being vaccinated, is to maintain proper hygienic practices, especially related to hand and food hygiene (de Graaf, van Beek, & Koopmans, 2016; Dennehy, 2000; FitzSimons, Hendrickx, Vorsters, & Van Damme, 2010).

Droplet transmitted viral infections exhibit the highest potential for rapid pandemic spread in large clusters of people (Heeney, 2006; Salathé et al., 2010). This is rather worrying as the percentage of the world population living in cities is estimated to have increased from 50% in 2008 to 70% in 2025 due to population growth, migration and opportune climate conditions. The spread of a possible pandemic in such urban areas may be hastened by the deterioration of sanitation often encountered in the case of overpopulation (Bell et al., 2009). Pandemic cases of diseases which have spread through droplets in recent years include: Severe Acute Respiratory Syndrome (SARS) in 2002-2003 (Tan, Li, Wang, Chen, & Wu, 2004), influenza H1N1 in 2009-2010 (Kanadiya & Sallar, 2011), Ebola 2013-2016 (Aylward et al., 2014; Dudas et al., 2017) and the ongoing COVID-19 pandemic (Ge, Yang, Xia, Fu, & Zhang, 2020; J. Wang & Du, 2020). For droplet transmitted infections, the steps that can be applied to reduce spreading of the epidemic include testing and detection, patient isolation, contact tracing and encouraging society to take specific actions including altering behavior regarding hygiene (Fung & Cairncross, 2006; Tan et al., 2004). Such behavior are of paramount importance when dealing with pandemics as they can often be transmitted before any symptoms occur (Wilder-Smith & Freedman, 2020). However, to achieve self-isolation or government mandated quarantine to prevent the spread, one has to be in possession of a sufficient supply of food. Despite hampered logistics and problems related to supply chain and storage, grocery stores have to be open because they represent public access to the purchase of food products that are necessary to survive. Nonetheless, one must also bear in mind that grocery stores are a place for possible transmission of many bacterial and viral pathogens (Bell et al., 2009; Dalton, New, & Health, 2006; Sinclair, Fahnestock, Feliz, Patel, & Perry, 2018), causing consumers to undertake various behavioral changes in their approach to shopping.

The latest pandemic case is the emergence of severe acute respiratory syndrome Coronavirus 2 (COVID-19). The scale of infection of the new virus is very serious in the public health sector and has a basic reproduction number of 2.24 – 3.58, whereas the SARS virus from 2002 was at the level of 1.20 – 1.32 (Lai, Shih, Ko, Tang, & Hsueh, 2020; Massad, Burattini, Lopez, & Coutinho, 2005; McCloskey et al., 2020). As of April 28$^{th}$ 2020, WHO confirmed 2,954,222 cases of COVID-19 with a total death toll of 202,597 victims (WHO, 2020a).

In the authors' analysis of literature on the behavioral changes caused by pandemics on individuals who frequent stationary stores for their shopping, one overarching phenomenon stands out: a distinct lack of research on this topic, despite there being extensive research on the economic havoc that can be caused by such mass behavioral changes. Such a dearth of research on the behavioral changes undertaken by consumers in response to the perceived threat of contagion during epidemics and pandemics is worrying, as it displays a lack of preparedness for the crisis that is to follow. This further indicates a more sophisticated need to measure what areas of consumer behaviors at stores are affected by the epidemic and therefore, the aim of the present study was an attempt to create such a measure.

## Literature review

The CDC (2020), along with many governments around the world, are encouraging citizens to practice social distancing and undergo quarantine as greatly as it possible in order to limit the spread and exposure of COVID-19. Such advisories and regulations disrupt normal routines, create anxiety and cause what (Forster & Tang, 2005) call a crisis of fear. The results of a survey examining the level of anxiety among students during a swine flu pandemic showed that 83.1% of respondents felt some kind of anxiety, with 5.1% feeling severely worried (Alnajjar, Attar, Farahat, & Althaqafi, 2016; Funk et al., 2010; Jones & Salathé, 2009). Sometimes, the chaos and pressure of information concerning the high mortality risk of pandemic causes misunderstandings and improper behavior, such as refusing vaccination or avoiding public health facilities. The results of a survey submitted by Jones and Salathé (2009) showed that the level of anxiety and preventive actions decreased with the perception of the seriousness of the outbreak and the high level of belief in avoiding infection. However, in most cases, it was been possible to introduce changes and increase public awareness related to personal and environmental hygiene as well as frequent disinfection (Balkhy, Abolfotouh, Al-Hathlool, & Al-Jumah, 2010; Jones & Salathé, 2009; Little et al., 2015; Zhang, Gu, & Kavanaugh, 2005). The higher level of anxiety, the greater was the implementation of preventive actions. The results of many studies indicated that more than half of the pandemic population was more likely to wash and disinfect their hands as an effective infection control intervention. Wearing face masks has also became common. In addition, house spaces are more frequently ventilated (Balkhy et al., 2010; Fleischman et al., 2011; Kanadiya & Sallar, 2011; Kantele et al., 2010; SteelFisher et al., 2012; Tan et al., 2004). Depending on nationality, respondents started to more often cough or sneeze into their elbow or shoulder (25 - 84%) and covered their mouth and nose with a tissue when coughing or sneezing (27-77%) (SteelFisher et al., 2012). During the COVID-19 epidemic in China, due to the high risk of infection, 84.7% of respondents spent 20-24 hours a day at home, and 53.8% of subjects rated the psychological impact of the outbreak as moderate or severe (C. Wang et al., 2020). A survey was conducted in several countries in response to the 2009 H1N1 pandemic. Results showed that respondents most often avoided places where larger group of people could gather, such as shopping centers or sports events (SteelFisher et al., 2012). According to other studies performed during epidemic, 70.7% of respondents limited all outdoor activities (Tan et al., 2004), and 64.8% believed avoiding crowded places is an effective preventive action (Kanadiya & Sallar, 2011).

The perceived fear during the pandemic, however, may be separate from the real threat posed by the disease in question, creating disproportionate behavioral changes among individuals. For example, during SARS epidemic in Hong Kong, 23% of respondents considered themselves "very likely" or "somewhat likely" to become infected with SARS at the peak of the epidemic, when the post-infection rate was only 0.0026% (Leung et al., 2004). Such exaggerated perceptions were also recorded in Taipei where 74% of surveyed respondents rated themselves as "4" or "5" on a 5-point scale to measure the chances of contracting SARS, leading to their death, when the actual mortality rate was 11% (Liu, Hammitt, Wang, & Tsou,

2005). Such individual subjective probability judgments about the risk of contraction cause mass avoidance of other individuals (Brahmbhatt & Dutta, 2008), initiating major economic disruptions (Noy & Shields, 2019).

A sector in which consumers maintained relatively persistent expenditures during times of an epidemic such as MERS in Korea, concerned groceries. Such an aspect cannot be postponed unlike discretionary spending (Jung, Park, Hong, & Hyun, 2016). However, the epidemic causes shopping behaviors of consumers to change. According to (Forster & Tang, 2005), the peak of SARS in Hong Kong drew an increasing number of consumers to online shopping for their staples such as canned goods and rice. Similar findings have been obtained by Jung et al. (2016) who discovered that the spread of MERS in Korea made consumers shift their spending to online portals and away from physical retail stores due to the risk of contagion. In a recent study focused on US household spending patterns amidst the COVID-19 pandemic, it was discovered that consumer spending dramatically increased in order to stockpile goods in anticipation of an inability to shop at retailers (Baker, Farrokhnia, Meyer, Pagel, & Yannelis, 2020). Another interesting observation from South Korea, provided by Nielson (2020a), was that the spread of COVID-19 is prompting consumers to reduce their visits to large supermarkets, and shift their shopping tendencies more towards neighborhood stores where they have little interaction with other consumers whilst only travelling short distances. The same author reports that a Korean family affair such as shopping has now become the responsibility of an adult member of the family in order to minimize the exposure of the remaining family members to potential threats. A survey conducted in Germany showed that 83.6% of respondents did not do shopping daily with at least 50% of the German population having stockpiled food to last for 10 to 11 days (Gerhold, 2020). Based on the analysis of this limited quantity of research related to consumer behavioral changes in response to epidemics, it is clear that there is a gap in research on how the fear of contagion and not budgetary limitations can impact consumer willingness to shop at stationery stores.

## Methodology and results

A number of activities were performed to develop a tool for measuring the dimensions of COVID-19 impact on the in-store behavior of consumers. This study was conducted in accordance with the guidelines for building scales (Churchill, 1979; Peter, 1981) and takes into account the proposal of Rossiter (2016) and its limitations (Bergkvist & Zhou, 2016; Lee & Cadogan, 2016; Salzberger, Sarstedt, & Diamantopoulos, 2016). The whole process involved 3 stages: qualitative inquiry, scale purification and scale validation, which were based on 3 studies: 1 qualitative and 2 quantitative. Qualitative data was used to prepare the first list of statements. On the basis of data from the first qualitative study, exploratory and then confirmatory factor analysis was conducted. The final part of the research included carrying out the study on a larger sample and on this basis, re-conducting confirmatory analysis. In quantitative research, the R programming environment and the GPA rotation, Psych, Lavaan packages were used as well as R-based programs: Jamovi and JASP.

*Qualitative inquiry*

Qualitative methodology was applied due to the exploratory nature of this research. The research team was particularly interested in developing a deeper understanding of how consumers behave at stores and choose the place of food purchase. Individual interviews were conducted in the study. The trial was semi-structured and included nearly 26 questions in total, except for the initial and demographic questions, which were intended to create an open atmosphere between the researcher and the participant. These questions were grouped into 4 areas: questions about the person doing grocery shopping, about the place of shopping, behavior at the store and questions regarding preferred products. Interviews were conducted remotely using the Zoom application and the entire conversations were recorded. The study was carried out among 20 respondents and each interview lasted on average of approx. 20-40 min. The subjects were diversified according to age, education, sex and place of residence.

*Qualitative results and item generation*

In the study, many differences were revealed in the approach to shopping during an epidemic emergency. All enquiries indicated that the epidemic has affected the way the respondents' shop. For some respondents, the change in behavior was due to the top-down restrictions rather than their own beliefs, while for some, these alterations related to changes in the place, time, frequency of purchase, and behavior in the store itself. Importantly, more attention was paid to the person or people shopping. In the case of some participants, not all the areas of possible epidemic impact and sense of threat were affected in the same way. What is more, opposing phrases appeared, e.g. regarding the size of the preferred store or its distance from place of residence. The form of the open interview allowed exploring motives for individual behaviors, which translated into the possibility of generating items. The original statement list, which was prepared for analysis of qualitative data, was linguistically modified. This modification included the elimination of negative forms in sentences, as well as complex and difficult formulations. Furthermore, a normative nature of the scale was adopted. The list of items was prepared as statements to which the respondent could refer. In connection with the implementation of the quantitative study among US residents, the original version of the questionnaire was prepared in English and verified by an American-English native speaker.

*Scale purification*

The basis for the first stage of quantitative research was a list of 62 items which were created on the basis of qualitative analysis. Data collection was preceded by a pilot study among 4 respondents to verify command clarity and eliminate possible restrictions. As a result, minor corrections were made. The respondents were recruited for the main study using the Amazon Mturk platform. The study involved 552 people, of whom 373 persons were included in the analysis on the basis of passing control questions that verified attention. The average age was 36 (SD = 12), 183 participants were women (49.06%), 189 men (50.67), 1 person did not answer the questions. The respondents were diverse due to education, income and professional status (Table 1).

Table 1. Description of the study group (study 2)

| Education | Frequency | Percentage | Cumulative Percentage |
|---|---|---|---|
| Bachelor's degree | 174 | 46.649 | 46.649 |
| Doctorate | 15 | 4.021 | 50.670 |
| High school degree or equivalent | 91 | 24.397 | 75.067 |
| Less than a high school diploma | 7 | 1.877 | 76.944 |
| Master's degree | 77 | 20.643 | 97.587 |
| Other | 9 | 2.413 | 100.000 |
| Missing | 0 | 0.000 | |
| Total | 373 | 100.000 | |

| Annual income | Frequency | Percentage | Cumulative Percentage |
|---|---|---|---|
| $20,000 – $29,999 | 55 | 14.745 | 14.745 |
| $30,000 – $39,999 | 48 | 12.869 | 27.614 |
| $40,000 – $49,999 | 25 | 6.702 | 34.316 |
| $50,000 – $59,999 | 34 | 9.115 | 43.432 |
| $60,000 – $69,999 | 20 | 5.362 | 48.794 |
| $70,000 – $79,999 | 28 | 7.507 | 56.300 |
| $80,000 – $89,999 | 11 | 2.949 | 59.249 |
| $90,000 ≥ | 44 | 11.796 | 71.046 |
| ≤ $19,999 | 108 | 28.954 | 100.000 |
| Missing | 0 | 0.000 | |
| Total | 373 | 100.000 | |

| Status | Frequency | Percentage | Cumulative Percentage |
|---|---|---|---|
| Full-time employment | 171 | 45.845 | 45.845 |
| Part-time employment | 53 | 14.209 | 60.054 |
| Retired | 17 | 4.558 | 64.611 |
| Self-employed | 43 | 11.528 | 76.139 |
| Student | 36 | 9.651 | 85.791 |
| Unable to work | 11 | 2.949 | 88.740 |
| Unemployed | 42 | 11.260 | 100.000 |
| No data | 0 | 0.000 | |
| Total | 373 | 100.000 | |

Based on the empirical material, an analysis of assumptions about the validity of factor analysis was performed. Bartlett's test of sphericity provided statistically significant results, and the overall Kaiser-Meyer-Olkin (KMO) measure of sampling adequacy reached .93. All items except for one statement (shop at the same store more frequently - .76) were above .8.

The next element of analysis was to determine the number of factors. Parallel analysis was carried out, based on which 8 factors were established. Then, exploratory factor analyzes with oblique rotation were proposed because of the presumed correlations among the construct's dimensions. Items that had a saturation below .4, and when they considered communities below .03, were eliminated from further research. In addition, items with a load of several factors were also excluded (Hair, Black, Babin, Anderson, & Tatham, 2009). Some of the items were characterized by high residual covariances, which was due to their synonymous nature, also being the basis for elimination. As a result of purification, a more complex pattern of consumer purchasing behaviors at the store emerged than that assumed at the stage of qualitative research. In addition, the procedure resulted in the elimination of some of the variables identified during in-depth interviews. An example of such an area is the use of personal protective equipment such as masks or gloves. This may be due to the determinants of certain behaviors through pre-defined rules that form the foundation of store security. Moreover, the introduction of restrictions on the number of customers or rules prevailing in the store eliminate the importance of store size. As a result, factors were identified relating to the shopping process, including the choice of place and time as well as to the selection and preferences of products.

*Factor characteristics*

The total number of 8 factors was identified. Contact Limitation (CL) factor include behaviors that are supposed to reduce the risk of coming across other people while shopping for food products. It should also be noted that the CDC (Burke, 2020) estimates the contamination risk from an infected individual to be approx. 0.45% for close contact with someone infected and 10.5% for household members. In light of this, the CL factor also includes limiting the indicator regarding the number of household members who do shopping. This indicator of shopping alone is also important, since family co-shopping is a strong socializing agent (Keller & Ruus, 2014) and changes in attitude related to co-shopping may affect inter-family relations.

   Food Supply Security (FSS) is the second identified factor. It involves behaviors related to purchase of non-easily perishable food products and their stockpiling. This considers indicators which include purchase of frozen, preserved or, in general, food products with long expiration dates. This may be caused by 2 main reasons: in the case that something happens to the global food chain, and to reduce the number of times an individual has to leave home for shopping and thus, risk getting infected. It should be noted that up until now, there has been no evidence that the COVID-19 outbreak has affected the global food safety and security at any rate (Fereidoon, 2020).

   Factor 3 identified as Food Product Familiarity (PF) involves the purchase of recognized/trusted products and brands. This includes the purchase of products which are familiar to the consumer but also the purchase of trusted food brands. This might be due to desire to shorten shopping time to a minimum or due to the attitude that the time of epidemic is no time for experimenting with unfamiliar food products.

   Shopping Time Optimization (STO) is a factor involving the reduction of time spent in a shop and is related to limiting the time an individual is exposed to contamination by strangers. This regards not only shopping quickly but also the reluctance to have any direct conversation with other individuals present in shop as indicated by the "move smoothly without stopping other" indicator. This is an important factor since shopping is often used as means to socialize and meet new people (Dawson, Bloch, & Ridgway, 1990).

   The Keeping Distance (KD) factor is indicated by attitudes related to maintaining a space between individuals within the shop, to ensure that even if there is an infected person present at the moment of shopping, distance will reduce the risk of contamination. This involves not only keeping one's distance in the line, but also directly in the shop when someone else is choosing products from the shelves. This factor is similar to the Contact Limit aspect, with the difference that it includes indicators of in-shop behaviors, while the Contact Limitation factor is more related to general avoidance of other individuals. Information about maintaining physical distance, usually of at least one meter between individuals, is widely spread by the media and governmental organizations (WHO, 2020b).

The next identified factor is Product Packaging (PP), which is related to attitudes towards packed and unpacked foods. These indicators are related to the most common food products that are often purchased unpacked, such as vegetables, bread and various ready-to-eat products, including unpacked nuts, confectionery, dried fruits, etc. The change in attitudes towards this factor during an epidemic may be relevant due to recent ambivalence in relation to food packages. On the one hand, there was a growing trend of so called zero-packaging, which included denouncing disposable plastic packages and promoted the purchase of unpacked food products (Beitzen-Heineke, Balta-Ozkan, & Reefke, 2017). On the other, the SARS-CoV-2 virus can remain infectious for 24-72 h on various surfaces (van Doremalen et al., 2020).

Another identified factor is the Number of Stores (NoS) that consumers use for shopping. This factor includes indicators related to how many shops a person chooses during shopping, but also if s/he avoids shopping at unfamiliar stores. This is related to time optimization since going to unfamiliar shops usually increases the time spent on shopping. The consumer then requires more time to find desired products.

The last identified factor is related to Personal Security (PS) during shopping and includes the implementation of protective gear such as gloves or masks and the use of disinfectants. One indicator of this factor is the use of disinfectants to sanitize handles after touching, for instance, freezer doors. This factor also regards the use of contactless payment methods as a protective measure, which is related to the warnings that physical money may be a source of virus transmission (WHO, 2020c)

*Scale validation (first data collection)*

Next, confirmatory factor analysis was performed. The indices show an acceptable fit to the data (RMSEA = .06, TLI = .94, CFI .94, SMRM = .07, RNI = .94) (Hu & Bentler, 1999). Factor loadings of all items are within the range from .62 to .96. In addition, discriminant validity of the 8-dimension scale was made. Analysis included correlation between constructs and verification whether the values were significantly below 1 (Bagozzi & Heatherton, 1994). The highest correlation between dimensions was 0.84 (between Contact Limitation and Product Optimization). The associated confidence interval was 0.74 to 0.93. Hence, discriminant validity was supported for all pairs of dimensions.

*Scale validation (second data collection)*

The authors re-examined the 8-dimensional scale of in-store consumer behaviors during a pandemic. Recruitment, as before, was carried out using Amazon Mturk among Americans who did not answer the previous questionnaire. The questionnaire containeds 5 questions verifying attention, the answers involved included duplicated reversed questions. The study included 584 responses from all 863 answers. The questionnaire included 362 women (62.16%) and 214 men (36.64%), and the average age of the respondents was almost 41 years (SD = 13.72). This question was not answered by 7 respondents. As in the first study, consumers were diversified based on education, income and employment status (Table 2). The study included 26 questions from the original 62.

Table 2. Description of the study group (study 3)

| Education | Frequency | Percentage | Cumulative percentage |
|---|---|---|---|
| Bachelor's degree | 271 | 46.40 | 46.40 |
| Doctorate | 16 | 2.74 | 49.14 |
| High school degree or equivalent | 169 | 28.94 | 78.08 |
| Less than a high school diploma | 2 | 0.34 | 78.42 |
| Master's degree | 99 | 16.95 | 95.38 |
| Other | 27 | 4.62 | 100.00 |
| Missing | 0 | 0.00 | |
| Total | 584 | 100.00 | |

| Annual income | Frequency | Percentage | Cumulative percentage |
|---|---|---|---|
| $20,000 – $29,999 | 79 | 13.53 | 13.53 |
| $30,000 – $39,999 | 66 | 11.30 | 24.83 |
| $40,000 – $49,999 | 60 | 10.27 | 35.10 |
| $50,000 – $59,999 | 61 | 10.45 | 45.55 |
| $60,000 – $69,999 | 43 | 7.36 | 52.91 |
| $70,000 – $79,999 | 41 | 7.02 | 59.93 |
| $80,000 – $89,999 | 28 | 4.79 | 64.73 |
| $90,000 ≥ | 101 | 17.29 | 82.02 |
| ≤ $19,999 | 105 | 17.98 | 100.00 |
| Missing | 0 | 0.00 | |
| Total | 584 | 100.00 | |

| Current status | Frequency | Percentage | Cumulative percentage |
|---|---|---|---|
| Full-time employment | 290 | 49.66 | 49.66 |
| Part-time employment | 92 | 15.75 | 65.41 |
| Retired | 40 | 6.85 | 72.26 |
| Self-employed | 59 | 10.10 | 82.36 |
| Student | 29 | 4.97 | 87.33 |
| Unable to work | 18 | 3.08 | 90.41 |
| Unemployed | 56 | 9.59 | 100.00 |
| Missing | 0 | 0.00 | |
| Total | 584 | 100.00 | |

Confirmatory factor analysis was carried out once more. Improvement was noted in all the analyzed measures (CFI = .96, TLI = .95, RNI = .96, GFI = .91, RMSEA = .05, SRMR = .05). The value of individual factors was above .73 except for 2 indicators: the use of contactless payment (.68) and stocking up on food items (.66) (Table 3). In these 2 cases, the achieved $R^2$ value also reached .44 and .47, respectively, and in other cases, from .54 to .94. Discriminant validity covering the (Fornell & Larcker, 1981) test indicated meeting the requirements for each of the 8 factors. Both Cronbach's A and Composite reliability demonstrated values above .87 with the exception of Personal Security. In this case, average variance exceeded the recommended .5, where in the case of other factors, these values ranged from .63 to .86 (Table 4).

Table 3. Results of Confirmatory Factor Analysis

| Factor | Indicator | z-value | p | Std. est. |
|---|---|---|---|---|
| CONTACT LIMITATION | doing shopping without accompanying people | 23.44 | < .001 | 0.82 |
| | limiting the number of household residents who do shopping | 21.03 | < .001 | 0.76 |
| | purchasing from a store that in which there are few customers at a time | 25.01 | < .001 | 0.85 |
| | doing shopping at times of low shopper traffic | 22.93 | < .001 | 0.81 |
| FOOD SUPPLY SECURITY | purchasing preserved food products | 22.92 | < .001 | 0.81 |
| | purchasing frozen food products | 23.42 | < .001 | 0.82 |
| | purchasing food products with long expiration dates | 26.28 | < .001 | 0.89 |
| | stocking up on food items | 17.35 | < .001 | 0.66 |

| PRODUCT FAMILIARITY | purchasing already known food products | 29.47 | < .001 | 0.93 |
| --- | --- | --- | --- | --- |
| | purchasing trusted food brands | 29.79 | < .001 | 0.94 |
| | choosing familiar products | 28.51 | < .001 | 0.91 |
| SHOPPING TIME OPTIMIZATION | limiting time spent in store | 27.89 | < .001 | 0.91 |
| | moving smoothly without stopping others | 22.76 | < .001 | 0.80 |
| | shopping quickly | 24.31 | < .001 | 0.84 |
| KEEPING DISTANCE | keeping a distance waiting in line | 30.73 | < .001 | 0.95 |
| | wait at a distance while someone else is choosing products | 27.56 | < .001 | 0.89 |
| | maintaining a distance while waiting in line | 28.34 | < .001 | 0.91 |
| PRODUCT PACKAGING | refraining from purchasing food without packaging | 27.85 | < .001 | 0.91 |
| | refraining from purchasing unpacked, ready-to-eat foods | 26.77 | < .001 | 0.89 |
| | limiting the purchase of unpackaged vegetables | 24.99 | < .001 | 0.85 |
| NUMBER OF STORES | using one grocery store for purchases | 28.59 | < .001 | 0.92 |
| | conducting shopping at one store | 31.28 | < .001 | 0.97 |
| | limiting the number of stores visited | 20.41 | < .001 | 0.73 |
| PERSONAL SECURITY | disinfecting handles after touching e.g. freezer doors | 18.70 | < .001 | 0.73 |
| | bringing hand disinfecting agent during shopping | 19.03 | < .001 | 0.74 |
| | using contactless payment | 17.09 | < .001 | 0.68 |

Table 4. Results of discriminant validity

| | CL | STO | PS | FSS | PF | KD | PP | NS | total |
| --- | --- | --- | --- | --- | --- | --- | --- | --- | --- |
| Cronbach's A | 0.87 | 0.89 | 0.76 | 0.87 | 0.95 | 0.94 | 0.92 | 0.9 | 0.94 |
| Composite reliability | 0.87 | 0.88 | 0.76 | 0.87 | 0.95 | 0.94 | 0.92 | 0.92 | 0.97 |
| Average variance extracted | 0.63 | 0.72 | 0.52 | 0.63 | 0.86 | 0.84 | 0.75 | 0.8 | 0.71 |

To study the nomological validity regarding dimensions concerning the impact of the epidemic state on consumer purchasing behaviors, the authors used a model including the identified factors along with the factor theoretically determining them, i.e. the fear of being infected by a virus at a store. The theoretical model includes the impact of in-Store Infection Threat (SIT) on the identified variables (Figure 1).

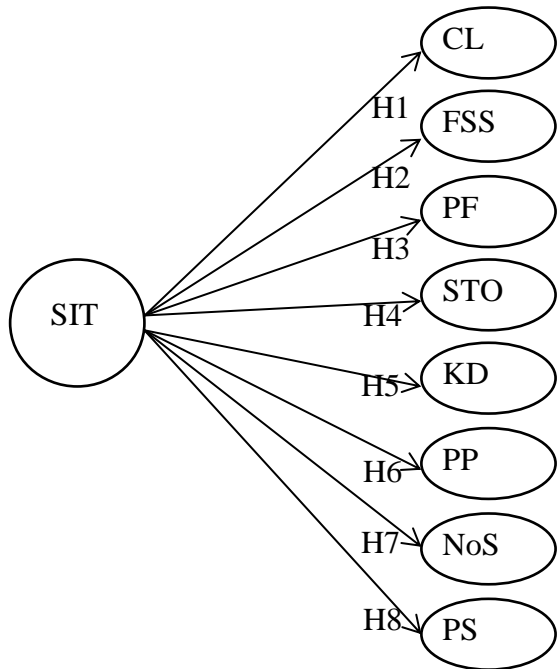

Figure 1. Research framework

A pandemic can cause individuals to undertake behavioral changes that are far from those truly required in accordance with pandemic severity. Such exaggerated perceptions of ones chances of being infected with virus was recorded in Taipei, where 74% of survey respondents rated their likelihood of contracting SARS, leading to their death, as very probable (Liu et al., 2005). The same was discovered in research from Hong Kong during the SARS epidemic, as 23% of respondents considered themselves "very likely" or "somewhat likely" to become infected when the post-infection rate was only 0.0026% (Leung et al., 2004). Such individual subjective probability judgments about the risk of contraction cause mass avoidance from other individuals (Brahmbhatt & Dutta, 2008), as was observed in South Korea during the spread of the COVID-19, where shopping has now become the responsibility of a single adult in the family (Nielson, 2020a). Such shopping, as per the same report, was also centered in neighborhood stores where the chance of interaction with other consumers is small, which is why we posit that:

- H1. Perceived in-store infection threat has positive impact on contact limitation.

Another change in behavior induced by an epidemic is re-assessment of the preferences and the importance of food attributes. Such changes was clearly observed during the SARS crisis in Hong Kong as there was a spike in the demand for rice, cooking oil, canned and consumable goods, frozen foods, cleaning products and toiletries (Forster & Tang, 2005). This increase in the purchase of items with long shelf-life such as powdered milk products, dried beans, canned meat, chickpeas, rice, tuna, black beans, biscuit mix, water and pasta was also evident during the current COVID-19 epidemic in the US (Nielson, 2020b). The same situation could be observed in Canada where the majority of items in consumer stockpiles consisted of canned, frozen, and fresh foods, along with toilet paper and hand sanitizers (Deloitte, 2020). With this past evidence the authors suggest the following hypotheses:

- H2. Perceived in-store infection threat has positive impact on food supply securing behaviors.
- H3. Perceived in-store infection threat has positive impact on the tendency to consume familiar products.

In a study by Balkhy et al. (2010) concerning statements and self-reported precautionary measures against H1N1 Influenza in Saudi Arabia, it was discovered that 51.6% of respondents preferred to stay at home during its duration. This aversion to conducting shopping in stores can find its justification in the research by Sadique et al. (2007), who discovered that venturing out to shops was considered the third riskiest setting in which one could acquire pandemic influenza, after places of entertainment and shops. In the same study, it was also concluded that 60% of respondents were partial towards doing only shopping that was considered essential. In the research conducted by Nielson (2020a) on South Korean consumers, analogous observations were noted, as the author were found that consumers reduced their visits to large supermarkets, and shifted more towards neighborhood stores where there is little interaction with other consumers. This interaction aversion behavior and dislike of instances where one can be exposed to the virus allow the authors to erect the following hypotheses:

- H4. Perceived in-store infection threat has positive impact on how consumers optimize their shopping time.
- H5. Perceived in-store infection threat has positive impact on in-store social distancing.
- H6. Perceived in-store infection threat has positive impact on the consumption of products without packaging.
- H7. Perceived in-store infection threat has positive impact on the number of stores frequented by the consumer.
- H8. Perceived in-store infection threat has positive impact on in-store behavioral changes taken to ensure one's personal safety.

In order to test the above hypotheses, in-store infection threat was measured using 5 items on a 7-point scale (There is a fear of becoming infected with the COVID-19 virus while shopping (SIT1), One can become infected with COVID-19 at the grocery store (SIT2),

Shopping during the COVID-19 epidemic is a risk to health (SIT3), There is a risk of infection with the COVID-19 virus while at the store (SIT4), When shopping, one is at risk of becoming infected with COVID-19 (SIT5)) among respondents participating in the second quantitative survey. Load values exceeded .79 (Table 5) and reached recommended values for the factor.

Table 5. Factor loadings of in-store infection threat (95% confidence interval)

| Factor | Indicator | Est. | SE | z-value | p | Lower | Upper | Std. est. |
|---|---|---|---|---|---|---|---|---|
| SIT | SIT1 | 1.13 | 0.05 | 23.26 | < .001 | 1.04 | 1.23 | 0.80 |
|  | SIT2 | 0.99 | 0.04 | 22.61 | < .001 | 0.90 | 1.07 | 0.79 |
|  | SIT3 | 1.22 | 0.04 | 27.98 | < .001 | 1.13 | 1.30 | 0.90 |
|  | SIT4 | 1.20 | 0.04 | 30.87 | < .001 | 1.13 | 1.28 | 0.95 |
|  | SIT5 | 1.22 | 0.04 | 29.47 | < .001 | 1.13 | 1.30 | 0.93 |

Cronbach's alpha = .94, CR=.94, AVE=.77.

A structural model was created to measure the impact of in-store infection threat on all identified dimensions of behavior in a store during an epidemic. The analyzes relied on a bootstrap procedure to ensure stability of the results across the whole sample. The model fit is very satisfactory ($\chi2$ / ddl = 2.51; TLI = 0.95; CFI = 0.96; GFI = .9, RMSEA = .05, SRMR = .05). The results of the analysis indicate that SIT positively affects all identified variables within the range from .21 for the PF factor to .54 for CL and the same for the STO factor (Table 6).

Table 6. Model parameter estimation: effects of in-store infection thereat

| Hypothesis | Latent factor | Indicator | B | SE | Z | Sig. | Beta | Result |
|---|---|---|---|---|---|---|---|---|
| H1 | CL | SIT | 0.65 | 0.10 | 6.76 | *** | 0.54 | Validated |
| H2 | STO | SIT | 0.63 | 0.09 | 6.86 | *** | 0.54 | Validated |
| H3 | FSS | SIT | 0.44 | 0.07 | 6.28 | *** | 0.41 | Validated |
| H4 | PP | SIT | 0.36 | 0.05 | 6.73 | *** | 0.34 | Validated |
| H5 | PF | SIT | 0.21 | 0.05 | 4.09 | *** | 0.21 | Validated |
| H6 | PS | SIT | 0.51 | 0.09 | 5.90 | *** | 0.46 | Validated |
| H7 | KD | SIT | 0.57 | 0.10 | 6.02 | *** | 0.50 | Validated |
| H8 | NoS | SIT | 0.44 | 0.06 | 7.34 | *** | 0.40 | Validated |

$P < 0.001$ ***

## Conclusions

This is the first study ever to design a scale of in-store behavior during an epidemic, which resulted in obtaining a validated scale with high confidence degree. The process of developing the scale included 1 qualitative and 2 quantitative methods. The resulting scale contains 26 items, with 8 dimensions of in-store behaviors. All identified factors correlate with the in-store infection threat which reiterates the importance of providing information that reveals the true scale of the pandemic and not leaving space for individuals to create subjective probability judgments. This is all the more important in order to support the April 2020 call from the WHO (2020b) to fight the so called "infodemic" that is flooding the average consumer. Since a great deal of this information is false or unreliable, it causes a serious problem for the consumers to recognize "true" recommendations. The fight against an infodemic such as the one experienced at present with COVID-19 and any future pandemics cannot be won without assessing consumers' attitudes and behaviors during an epidemic. The scale that has been developed in this study can be useful for assessing consumer compliance with official recommendations and may be a valuable tool in targeting gaps in consumer education and knowledge.

      The authors expect that the provision of information revealing the true severity of the pandemic to the general public will also reduce panic-buying associated with the onsets of pandemics, reducing the strain on supply chains. Doing so will allow citizens to go about their shopping in a rational manner, without the worry of any impending inability to do their shopping to feed their families. This research also has important implications for stationary store outlets as they could initiate changes in store layout to accommodate any pandemic induced precautionary behaviors from their consumers. Other changes that stationary shops could undertake to accommodate pandemic-induced consumer behavioral changes include the provision of disinfections, disposable gloves, covering fresh produce such as bread, fruits and vegetables with protective covering, encouraging customers to make payments by cards, limiting the number of patrons in the store, marking distances at which consumers waiting in line should adhere to and increasing the stock of staple goods with long shelf-life.

      The authors hope that the implications of this research provide governments and policymakers with an understanding of how the timely provision of correct information using the right mediums can prevent consumers from making their own probability judgements about the threat of infection, which, in turn, leads to a climate of distrust, panic-purchasing and mass avoidance of stationary stores. This is all the more important as urbanization is at an all-time high with a majority of consumers depending on supermarkets for their shopping needs. This research, as previously mentioned, may also prove to be useful for manufacturers and stationery stores to adjust their supply of products to the demand shocks that are to be expected with the onset of a crisis such as the COVID-19 pandemic.

**Limitations and future research**

Although the study was designed in a way to be as precise as possible, some limitations exist. The main limitation is that the study was performed on only consumers from the USA and although the study included large number of participants with different metrics, it may still be difficult to apply this scale to consumers from countries with a different cultural background. Therefore, in future research, the questionnaire should be translated into different languages and performed among consumers from other countries affected by the epidemic. Moreover, the questionnaire was performed during the outbreak of the COVID-19 pandemic which limits the possibility of comparing the results for in-shop behaviors with a time from before the epidemic. Moreover, some responses may have been affected by government-imposed restrictions.

The data and findings obtained in this study raise several interesting avenues for future research. The first is honing in on the demographics of the research sample in order to identify whether factors such as education, income and employment status reveal discrepancies in the degree to which the threat of the virus is considered. Further research into this could then disclose their correlations with the 8 factors proposed in this article. Another area of research that could prove to be interesting, in order to discover the information medium upon which consumers' subjective opinions about the probability of contracting the virus are founded, is the amalgamation of data on where consumers receive their information on pandemics and its spread with a model such as the one presented by the authors of this study. There is also the possibility of extending this model to find out whether variables such as preexisting health conditions, the number of family members and the possibility of remote work has impact on how the threat of the virus is perceived and how behavioral changes influence in-store shopping.